\documentclass[12pt]{iopart}
\usepackage{epsfig}
\usepackage{graphicx}
\usepackage{dcolumn}
\usepackage{bm}
\usepackage{iopams}

\begin{document}

\title[Scaling and Universality in two dimensions]{Scaling and universality in two dimensions: three-body bound states with short-ranged interactions}

\author{F F Bellotti$^{1}$, T Frederico$^{1}$, M T Yamashita$^{2}$, D V Fedorov$^{3}$, A S Jensen$^{3}$ and N T Zinner$^{3}$}

\address{
  $^{1}$Instituto Tecnol\'ogico de Aeron\'autica, DCTA,  12.228-900 S\~ao Jos\'e dos Campos, SP, Brazil\\
  $^{2}$Instituto de F\'\i sica Te\'orica, UNESP - Univ Estadual Paulista, C.P. 70532-2, CEP 01156-970, S\~ao Paulo, SP, Brazil\\
  $^{3}$Department of Physics and Astronomy - Aarhus University, Ny Munkegade, bygn. 1520, DK-8000 \AA rhus C, Denmark
}

\date{\today }

\begin{abstract}
The momentum space zero-range model is used to investigate universal
properties of three interacting particles confined to two dimensions. The
pertinent equations are first formulated for a system of two identical and
one distinct particle and the two different two-body subsystems are characterized
by two-body energies and masses.
The three-body energy in units of one of the two-body energies is a universal
function of the other two-body energy and the mass ratio. We derive
convenient analytical formulae for calculations of the three-body energy as
function of these two independent parameters and exhibit the results
as universal curves. In particular, we show that the
three-body system can have any number of stable bound states. When the mass
ratio of the distinct to identical particles is greater than $0.22$ we find
that at most two stable bound states exist, while for two heavy and one light mass
an increasing number of bound states is possible. The specific number
of stable bound states depends on the ratio of two-body bound state energies
and on the mass ratio and we map out an energy-mass phase-diagram of the number
of stable bound states. Realizable systems
of both fermions and bosons are discussed in this framework.
\end{abstract}

\maketitle

\section{Introduction}
Quantum mechanics in two-dimensional (2D) systems differs quite
markedly from the three-dimensional (3D) case in many aspects.
A particular example is the centrifugal barrier
which is zero or positive in 3D, whereas in 2D
the $s$-wave barrier is in fact negative. This implies
that two particles in 2D are at the threshold of binding
even when they do not interact, i.e. an infinitesimal attraction
will produce a bound state \cite{nie97,nie99,nie01}. This is seen in the famous Landau
criterion which says that potentials with negative volume integral
will produce a bound state for any value of the strength
\cite{landau1977}. In fact, even when the volume integral is exactly
zero a bound state is still present
\cite{simon1976,jeremy2010,klawunn2010,baranov2011,artem2011a,artem2011b}.
These results are in sharp contrast to the 3D arena
where it is well-known that a finite amount of attraction is
required to produce bound states.

A natural question  to ask is how these dimensional differences
influence few-body states with more than two particles. A direction
of research has therefore been to derive and establish conditions
for the occurrence of properties of such systems. Most prominent
among the findings is arguably the Efimov effect \cite{efi70} which
has achieved a unique position as a mathematical anomaly of three
interacting particles in 3D. The so-called Efimov
states generally correspond to excited states that reside at large
distance and owe their properties to large two-body scattering
lengths. The relative properties, however, do not even depend on the
scattering length but only on the mass ratios of the constituent
particles which determine for instance the scaling of energy and
size from one state to the next. These scalings are completely
independent of the short-distance details of the inter-particle
potentials.

A number of universal relations have been found and studied in 3D.
The classical examples are correlations between two
observables where a one-dimensional relation emerges when the
potentials are varied. The Phillips plot is one curve for different
potentials when the neutron-deuteron scattering length is plotted versus
the triton binding energy \cite{phi68}, the Coester line of
saturation density versus binding energy per nucleon for nuclear
matter \cite{coe70}, and the Tjon line is the correlation between
binding energies of three and four nucleons \cite{tjo75}. Other much
more abundant 3D systems with very little model
dependence have been found and baptized halo systems \cite{jen04}.
They were at first primarily found as states in nuclei but quickly
they were also search for in molecules, and in particular as helium
dimers and trimers \cite{nie98}. Universal relations have been
applied to check for Efimov excited states in nuclear halo systems
\cite{amorim97} and also in molecules \cite{delfino00}.

The study of the Efimov  effect has seen a revival in recent years
within the context of ultracold atomic gases after its initial
observation in a gas of Cesium atoms in 2006 \cite{kraemer2006}.
This was followed by a number of experiments using various atomic
species and a new subfield dubbed 'Efimov Physics' has since
developed (see the recent review in Ref.~\cite{ferlaino2010}). Some
of the latest experiments have demonstrated heteronuclear Efimov
states \cite{barontini2009} and also the presence of some intriguing
four-body states \cite{platter2004,yamashita2006,stecher2009}
tied to the Efimov trimers \cite{pollock2009,ferlaino2009}.

Common to all the studies and experiments mentioned thus far is
that they are 3D. If one considers instead a
2D setup, then the structure of the three-body states
is very different. In the case of three equal mass particles the
tower of states close to threshold is absent and one finds just two
bound states in the universal limit where the particles 
reside outside the range of the two-body potential \cite{nie97,tjon79}. 
Here we
investigate a system of three particles where two are assumed
identical as function of the two-body interaction energies and the
mass ratios. As we will demonstrate, one can potentially have any
number of stable bound states when the parameters of the system are
varied. We therefore find a much more varied picture of
2D three-body systems in the universal limit than has
previously been considered.

Two-dimensional physics is important for a variety of problems such
as high-temperature superconductors, localization of atoms on
surfaces, in semiconducting microcavities, and for carbon nanotubes
and organic interfaces. One particular interesting example is the
two-component Fermi gas with attractive interactions in 2D. Here the
pairing instability is caused by the presence of a 
two-body bound $s$-wave state
\cite{randeria1989,schmitt1989,dreschler1992}. The role of
three-body states, however, remains elusive. In the ultracold atomic
gas community there is great interest in producing quantum
degenerate gases in low dimensions to investigate some of the basic
features of a 2D environment. Early experiments succeeded in
producing quasi-2D samples of $^{133}$Cs
\cite{cesium2D1,cesium2D2,cesium2D3}, $^{23}$Na \cite{sodium2D}, and
$^{87}$Rb \cite{rubium2D}. Mixtures of $^{40}$K and $^{87}$Rb have
been used to produce 2D gases \cite{modugno2003,gunter2005}, and
more recently samples with two-component gases of $^{6}$Li
\cite{marti2010,dyke2011} and $^{40}$K \cite{frohlich2011} have been
studied. Heteronuclear molecules $^{40}$K$^{87}$Rb have also been trapped
in a layered stack of quasi-2D pancakes \cite{miranda2011}.

Quasi-2D system are abundant in ultracold atom physics and we
thus expect that our predictions can be studied in a variety of
these setups. In a real experiment the confinement to 2D is
typically done using an optical lattice. This introduces a
transverse energy scale, $\hbar\omega_0$. Below this scale the
physics is effectively 2D while it becomes 3D at or above
$\hbar\omega_0$. In this first study we will assume a strict 2D
setup, which amounts to assuming that we are always working far
below $\hbar\omega_0$. In some of the limiting cases considered here
this assumption breaks down and in those cases the trap must be
explicitly included to make predictions for real experiments.
However, in the weak-coupling limit where the two-body bound state
energies of all three subsystems is small we expect our results to
apply directly. The two-body energy must, however, be modified to
include the trap scale in the manner outlined in
Ref.~\cite{bloch2008}. In addition, confinement-induced resonances 
occur in low-dimensional systems \cite{haller2010} and have to 
be taken into account when considering the possible few-body
bound states.
We expect that techniques for detection of
three-body states and resonances in 3D will also be useful in the
quasi-2D case, i.e. identification of sharp features in the loss
rates \cite{ferlaino2010} or radio-frequency association
\cite{lompe2010,nakajima2011}.

The three-body states we  consider here consist of two identical
particles and a third distinct particle. This can be a system
consisting of two identical bosons and a third particle, or two
fermions with at least two internal degrees of freedom (typically
hyperfine projections in the case of ultracold alkali gases). As we
consider the $s$-wave channel only in our equations, this means that
in the case of two identical fermions, the interaction is
necessarily zero. While there could be interaction in higher partial
waves, we neglect these in the present study as they are usually
much smaller than the $s$-wave interactions.

\begin{figure}[htb!]
\centering
\epsfig{file=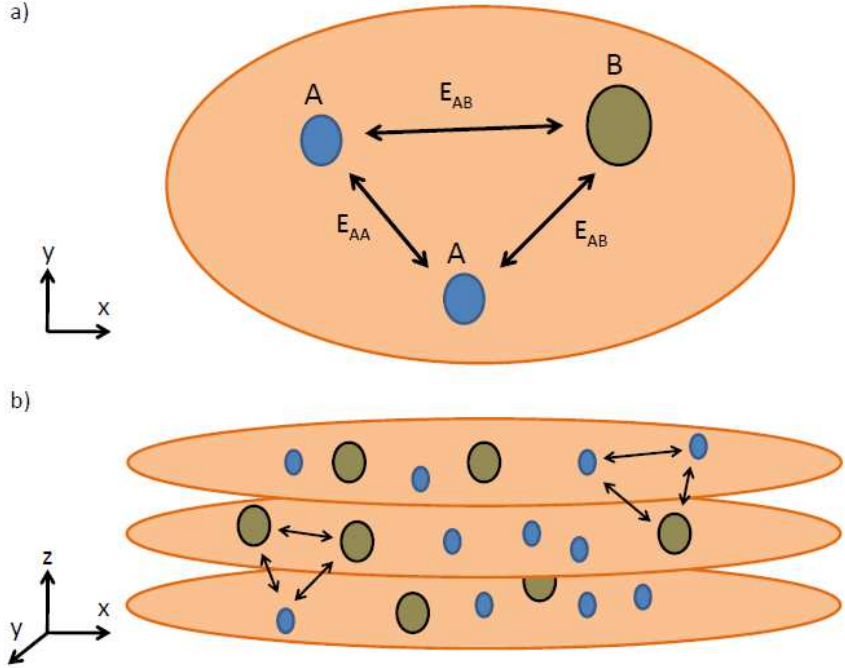,width=0.75\textwidth,angle=0}
\caption{Schematic illustration of the generic setup considered.
a) A three-body system in a plane (depicted more realistically
as a pancake with ellipsoidal shape to indicate potential trap 
deformation) with two identical A particles and one B particle
(top view). The picture indicates that the mass of B is
larger than that of A. Two-body interaction energies, $E_{AA}$
and $E_{AB}$, and coordinate axis are also shown.
b) A multi-layer setup containing multiple A and B particles
(side view). In a system with long-range interactions
(like dipole-dipole forces) three-body states can be formed from
particles in the same plane and in different planes. Two kinds 
of three-body systems with particles in adjancent layers 
are indicated by arrows. Coordinate system
is shown on the left.}
\label{schematic}
\end{figure}

We illustrate the generic  system under consideration in
figure~\ref{schematic}a) as a plane with two identical A particles
and one B particle. The plane is drawn as a pancake which is the
typical experimental situation. In the picture we assume that the
mass of A is smaller than that of B but we will consider both cases
below. In figure~\ref{schematic}b) we show another interesting setup
for which our results are relevant, a multi-layer stack of pancakes
with particles of both kind in each of the layers. In addition to
the in-plane three-body states that are possible, a long-range force
that acts across the layers can also provide binding. 
Our results are then relevant in
the universal limit which applies for ultracold polar molecules with
small dipole moments. This
has been discussed in one- \cite{wunsch2011} and three-dimensional
systems very recently \cite{greene2011}, and for the 2D bi- and
multi-layers in the weak
\cite{jeremy2010,klawunn2010,baranov2011,artem2011a,artem2011b} and
strongly coupled cases \cite{jeremy2011}. This 2D multi-layer with
dipolar particles has been experimentally realized as mentioned
above \cite{miranda2011}.

Occurrence conditions and properties of model independent or
universal states, like in Efimov or halo physics \cite{fbs11}, are
studied using zero-range models where the properties
of the systems arise from distances larger than the interaction
ranges \cite{adh88,adh93}. The behavior is important as a measure
against such asymptotic properties, but clearly also directly if
these properties can be realized in Nature or in laboratories. Both
these roles now appear more and more in 2D physics. The
purpose of the present paper is to provide universal energy
relations for three-body systems in 2D. These relations
are completely general and specific applications only require
appropriate two-body input parameters.

The structure of the paper is as follows. The introduction is followed
by a sketch in section II of the method. Equipped with the appropriate
equations the numerical investigations follow in section III. The stability
of ground and excited states are studied in section IV. Along the way we
reformulate the integral equation to apply to various threshold properties.
In section V we briefly sum up and formulate the conclusions.

\section{Formalism and Notation}
We consider a two-dimensional $AAB$ system assuming that the interactions
depend only on relative distances. This can be a three-body system in
a single plane as in figure~\ref{schematic}a) or three particles in
two or three layers when long-range interactions are present as in
figure~\ref{schematic}b).
In both situations, the three-particle dynamics effectively happens
in a single plane. We use zero-range
interactions as we are interesting in the for model-independent 
universal limit. This simplifies the
formulation of the Faddeev equation for the three-body bound state as the
contact interaction is separable. In the case of long-range interaction
across different layers, the use of zero-range interactions assumes that
the low-energy properties of the true long-range interaction can be described by
an effective interaction of zero-range similarly to the van der Waals
interaction for neutral atoms in 3D. This is true for potentials that 
go to zero at infinity faster than $1/r^2$ in 2D \cite{verhaar1985}, 
which includes the dipolar $1/r^3$ interaction.

The 2D Hamiltonian for the three-particle $AAB$ system with a
pairwise potential is
 \begin{equation} H=H_0+V_{AA}+V_{AB}+V_{AB}
\end{equation}
where the two $A$ particles are assumed to be identical bosons. The kinetic energy
operator is
\begin{equation}
H_0=\frac{\mathbf{q}_A^2}{2\mu_{A,AB}}
+\frac{\mathbf{p}^2_A}{2\mu_{AB}}
=\frac{\mathbf{q}_B^2}{2\mu_{B,AA}}
+\frac{\mathbf{p}^2_B}{2\mu_{AA}}
  =\frac{\mathbf{q}_{A^\prime}^2}{2\mu_{A,AB}}
+\frac{\mathbf{p}^2_{A^\prime}}{2\mu_{AB}},
\end{equation}
where we use Jacobi relative momenta given in terms of
rest frame momenta,  $\mathbf{k}_i$ with $i=\alpha,\beta,\gamma$,
as
\begin{equation}
\mathbf{q}_\gamma=\mathbf{k}_\gamma \,\, \textrm{and} \, \, 
\mathbf{p}_\gamma=\mu_{\alpha\beta}\left(\frac{\mathbf{k}_\alpha}{m_\alpha}-\frac{\mathbf{k}_{\beta}}{m_\beta}\right)
\label{jacobi}
\end{equation}
where $(\alpha,\beta,\gamma)$ is the cyclic permutations of the
particles $(A,A^\prime,B)$ with masses $m_A$ and $m_B$. The
reduced masses are $\mu_{\alpha\beta}=\frac{m_\alpha
m_\beta}{m_\alpha+m_\beta}$ and
$\mu_{\gamma,\alpha\beta}=\frac{m_\gamma(m_\alpha
+m_\beta)}{m_\alpha+m_\beta+m_\gamma}$. The contact potential in
operator form is
\begin{equation}
V_{\alpha\beta}=\lambda_{\alpha\beta}|\chi_{\alpha\beta}\rangle\langle\chi_{\alpha\beta}|,
\end{equation}
where the form factor $\chi_{\alpha\beta}(\mathbf{p}_\gamma)=1$,
depends only on the relative momentum of the pair.

The two-body T-matrix for negative energies and zero-range
potentials for $AA$ and $AB$ subsystems is
\begin{equation}
T_{A\gamma}(E)=|\chi_{A\gamma}\rangle\tau_{A\gamma}(E)\langle\chi_{A\gamma}|,
\label{T}
\end{equation}
where, the matrix element of the 2D transition matrices are given by
(see e.g. \cite{adh88} for the case of identical particles)
\begin{eqnarray}
\tau_{A\gamma}(E)=\left[ -4\pi \frac{m_{A}m_{\gamma}}{%
m_{A}+m_{\gamma}}\ln \left( \sqrt{\frac{E}{E_{A\gamma}}}\right)
\right] ^{-1} \ , \label{tau}
\end{eqnarray}
where $\gamma=A$ or $B$ and $E_{A\gamma}$ is the energy of the $AA$
and $AB$ two-body bound states. We adopt units such that $\hbar=1$.
The singularity of the two-body scattering equation in 2D is
subtracted by fixing the pole at the two-body bound state (see e.g.
\cite{adh1995}). To find the bound $AAB$ system we concentrate on
the negative energy region but note that for positive energy
scattering, the analytic extension can be easily performed in
(\ref{tau}). We note that effective range corrections have recently been 
discussed for three identical bosons \cite{helfrich2011} and
will in general shift the binding energies. Such corrections can
be included in the current formalism and can presumably be 
tuned in experiment by chosing different Feshbach resonances and
trapping frequencies. However, since
we are mainly interested in 
the overall structures (such as the number of bound states for 
given masses and interaction ratios) we neglect such corrections
in this study. 

The bound state wave function $|\Psi_{AAB}\rangle$ decomposed in
terms of the Faddeev components is
\begin{equation}
|\Psi_{AAB}\rangle=|\Psi_{A}\rangle+|\Psi_{A^\prime}\rangle+|\Psi_{B}\rangle
\end{equation}
where
$|\Psi_{\gamma}\rangle=G_0(E)V_{\alpha\beta}|\Psi_{AAB}\rangle$ with
the resolvent $G_0(E)=(E-H_0)^{-1}$ which is nonsingular for bound
states. The Faddeev equations are written in terms of the transition
matrix, which is well defined for the zero range potential and given
by (\ref{T}). We then have
\begin{equation}
|\Psi_{\gamma}\rangle=G_0(E)T_{\alpha\beta}\left(E-\frac{\mathbf{q}_\gamma^2}{2\mu_{\gamma,\alpha\beta}}\right)
\left(|\Psi_{\alpha}\rangle+|\Psi_{\beta}\rangle\right) \ ,
\label{FE}
\end{equation}
which simplifies through the separability of the potential to give
\begin{equation}
|\Psi_{\gamma}\rangle=G_0(E)|\chi_{\alpha\beta}\rangle|f_\gamma\rangle
\ , \label{vertf}
\end{equation}
where $\langle
\mathbf{q}_\gamma|f_\gamma\rangle=\lambda_{\alpha\beta}\langle
\mathbf{q}_\gamma,\chi_{\alpha\beta}|\Psi_{AAB}\rangle$ is the
vertex function of the Faddeev component of the wave function.

The coupled set of homogeneous integral equations for the Faddeev
components of the vertices of the wave function for the bound $AAB$
system  follows from (\ref{FE}) and (\ref{vertf}) and gives (see e.g.
\cite{adh88} and \cite{adh93} for three identical bosons)
\begin{eqnarray}
f_{B}\left( \mathbf{q}\right) = &2\tau_{AA}\left(E_3-\frac{
m_{B}+2m_{A}}{4m_{A}m_{B}}\mathbf{q}^{2}\right)\times &\nonumber\\
&\int
d^2k\frac{f_{A}\left( \mathbf{k}\right) }{E_{3}-\frac{
m_{A}+m_{B}}{2m_{A}m_{B}}\mathbf{q}^{2}-\mathbf{k}^{2}-\mathbf{k}\cdot
\mathbf{q}},& \label{eq.020}
\end{eqnarray}
\begin{eqnarray}
f_{A}\left( \mathbf{q}\right)
=&\tau_{AB}\left(E_3-\frac{m_{B}+2m_{A}}{2m_{A}\left(
m_{B}+m_{A}\right) }\mathbf{q}^{2}\right)\times &\nonumber\\
&\left[ \int d^2k\frac{f_{B}\left( \mathbf{k}
\right)
}{E_{3}-\mathbf{q}^{2}-\frac{m_{A}+m_{B}}{2m_{A}m_{B}}\mathbf{k}^{2}
-\mathbf{k}\cdot \mathbf{q}} +\right. &\nonumber \\
&\left.
\int d^2k\frac{%
f_{A}\left( \mathbf{k}\right) }{E_{3}-\frac{m_{A}+m_{B}}{
2m_{A}m_{B}}\left( \mathbf{q}^{2}+\mathbf{k}^{2}\right) -\frac{m_{A}}{m_{B}}
\mathbf{k}\cdot \mathbf{q}}\right],&   \label{eq.021}
\end{eqnarray}
where the $AA$ and $AB$ transition amplitudes are calculated for the
energies of the corresponding subsystems through (\ref{FE}). 
We note that the coupled set of
homogeneous equations described above gives the 3D equations for a
bound system $AAB$ when the transition amplitudes and momentum
volume are conveniently substituted by their 3D forms
\cite{amorim97}.

The $AAB$ wave function for the zero-range interaction is a solution
of the 2D free Schr\"odinger equation except when the
particles overlap. In momentum space, the wave function is built from
the spectator functions, $f_{A}\left( \mathbf{q}_{A}\right)$ and
$f_{B}\left( \mathbf{q}_{B}\right)$, which are solutions of the coupled
equations (\ref{eq.020}) and (\ref{eq.021}). The relative Jacobi
momentum of particle $A$ with respect to the center of mass of $AB$ is
given by $\mathbf{q}_{A}$ and for $B$ with respect to $AA$ by
$\mathbf{q}_{B}$. The $AAB$ wave function becomes
\begin{eqnarray}
\Psi_{AAB}\left(\mathbf{q}_{A},\mathbf{p}_{A}\right)=\frac{f_{A}\left(
\mathbf{q}_{A}\right)+f_{A}\left(
\mathbf{q}_{A^\prime}\right)+f_{B}\left( \mathbf{q}_{B}\right)}{E_{3}-\frac{2m_{A}+m_{B}}{%
2m_{A}(m_A+m_{B})}\mathbf{q}_{A}^{2}-\frac{m_A+m_B}{2m_Am_B}\mathbf{p}_{A}^{2}
} \ ,
\end{eqnarray}
where the relative Jacobi momenta $\mathbf{q}_{B}$ and
$\mathbf{q}_{A^\prime}$  re combinations of the pair
$(\mathbf{q}_{A},\mathbf{p}_{A})$ with $\mathbf{p}_{A}$ the relative
momentum between $A$ and $B$ defined in (\ref{jacobi}).

\section{Universal properties}
We want to find the $s$-wave three-body binding energy, $E_{3}$, for
a system of one distinct $B$ particle
and two identical $A$
particles. This yields $E_{3}$ as function of the two masses,
$(m_{A},m_{B})$, and the two-body binding energies,
$(E_{AA},E_{AB})$. These four quantities are the only unknown
parameters in the set of equations (\ref{eq.020})  and
(\ref{eq.021}) which determine $E_{3}$. Therefore
$E_{3}$ must be a function of these four parameters. By using one of
the binding energies, $E_{AB}$, as the unit of energy and
$m_{A}$ as the mass unit, we see from (\ref{eq.020}) and
(\ref{eq.021}) that the scaled three-body energy, $\epsilon_{3}$
can be expressed in terms of only two independent dimensionless
variables, that is
\begin{equation}
\epsilon_{3}\equiv \frac{E_{3}}{E_{AB}}={\cal F}\left( \frac{E_{AA}}{E_{AB}},\frac{%
m_{B}}{m_{A}}\right) \equiv {\cal F}_{n}(\epsilon_{AA},m)  \label{eq.023},
\end{equation}
where we used the definitions
\begin{equation}
m=\frac{m_{B}}{m_{A}}\,\, ,\,\,\epsilon_{3}=\frac{E_{3}}{E_{AB}}, \,\textrm{and}\,
\epsilon_{AA}=\frac{E_{AA}}{E_{AB}}.  \label{eq.025}
\end{equation}%
The universal functions, ${\cal F}_{n}$, are distinctly different for
ground and excited states as indicated by the discrete subscript
$n$. We shall predominantly focus on the ground state ($n=0$) but
also extract information about the number of bound states for a
given set of parameters. When the $A$
particles are identical fermions, the interaction in $s$-waves is
zero. The limit $\epsilon_{AA}\to 0$ is thus the relevant one for identical
fermions in the zero-range model employed here. We note that this
ignores interactions between the fermions in the $p$-wave channel
which are usually always negligible in comparison to the 
interactions in the $s$-wave channel between 
non-identical $A$ and $B$ particles.

Using mass unit $m_{A}=1$ and choosing the binding energy
$|E_{AB}|=1$, which is equivalent to rescaling all momenta by
$\mathbf{q}\;(\mathbf{k})\to
\sqrt{|E_{AB}|}\mathbf{q}\;(\mathbf{k})$  in the set of coupled
homogeneous integral equations (\ref{eq.020}) and
(\ref{eq.021}), one obtains
\begin{eqnarray}
f_{B}\left( \mathbf{q}\right) =&\left[ \pi \ln \left( \sqrt{\frac{
\frac{m+2}{4m}\mathbf{q}^{2}+\epsilon_{3}}{\epsilon_{AA}}}\right)
\right] ^{-1}\times&\nonumber\\
&\int d^{2}k\frac{f_{A}\left( \mathbf{k}\right) }{
\epsilon_{3}+\frac{m+1}{2m}\mathbf{q}^{2}+\mathbf{k}^{2}+\mathbf{k}\cdot
\mathbf{q}},&  \label{eq.026}
\end{eqnarray}

\begin{eqnarray}
f_{A}\left( \mathbf{q}\right) =&\left[ 4\pi \frac{m}{m+1}\ln \left(
\sqrt{\frac{m+2}{2\left( m+1\right) }\mathbf{q}^{2}+\epsilon_{3}}
\right) \right] ^{-1}\times & \nonumber\\
&\left[ \int d^{2}k\frac{f_{B}\left(
\mathbf{k}\right)} {\epsilon_{3}+\mathbf{q}^{2}+\frac{m+1}{2m}
\mathbf{k}^{2}+\mathbf{k}\cdot
\mathbf{q}}+\right.& \nonumber \\ & \left.\int
d^{2}k\frac{f_{A}\left( \mathbf{k}\right) }{\epsilon_{3}
+\frac{m+1}{2m}\left(\mathbf{q}^{2}+\mathbf{k}^{2}\right)
+\frac{1}{m}\mathbf{k}\cdot \mathbf{q}}\right].& \label{eq.027}
\end{eqnarray}
The coupled integral equations, (\ref{eq.026}) and (\ref{eq.027}),
give the scaled three-body energies,
$\epsilon_{3}$ as functions, ${\cal F}_{n}(\epsilon_{AA},m)$, of the two independent
parameters. We emphasize that $E_{AA}$,
$E_{AB}$, and $E_{3}$ are binding energies while $\epsilon_{3}$ and
$\epsilon_{AA}$ are ratios of binding energies and
will be referred to as scaled three- and two-body energy, respectively. An
analytical solution is not available and we shall instead
investigate the functions ${\cal F}_{n}$ by numerical means. We first
concentrate on ${\cal F}_{0}$ and its dependence on energy and mass.

\subsection{Dependence on two-body energy}
The solutions are firmly established for the completely symmetric case of
$\epsilon_{AA}=1$ and $m=1$, where all masses are equal and all
pairs have the same binding energy. Then two, and only two, bound
three-body states exist with energies given by
\cite{nie97,nie99,nie01,tjon79}
\begin{equation}
\epsilon_{3}^{\left( 0\right) }=16.52\;\;\textrm{and}\;\;\epsilon_{3}^{\left(1\right) }=1.27,
\label{eq.030}
\end{equation}%
where the superscript denote ground and first excited state,
respectively. These results have been confirmed by several studies
using different numerical techniques \cite{hammer2004,blume2005,brodsky2006,kartav2006}.
The two-body binding energies are
$E_{AB}=E_{AA}=E_{2}=-4\hbar ^{2}/(\mu a^{2})\exp (-2\gamma )$, where $a$ is
the scattering length, $\mu $ is the two-body reduced mass and
$\gamma =0.57721$ is Euler's constant.

\begin{figure}[htb!]
\centering
\epsfig{file=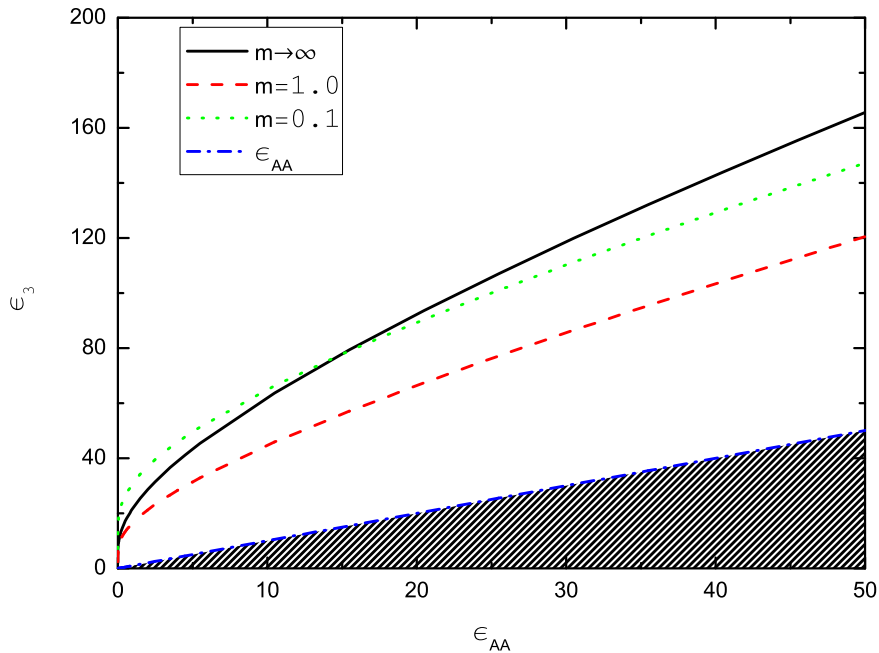,width=0.75\textwidth,angle=0}
\caption{Scaled three-body binding energy, $\epsilon_{3}={\cal F}_0(\epsilon_{AA},m)$, as
  function of the scaled two-body energy $\left( \epsilon_{AA}\right)$ for
  the ground state with $m=0.1$, $m=1.0$, and
  $m\rightarrow\infty$. States with energies above the $\epsilon_{3}=\epsilon_{AA}$ line
  are stable.  The highest-lying curve for large $\epsilon_{AA}$ corresponding to infinite mass is the
  function ${\cal A}_0(\epsilon_{AA})$ in (\ref{eq.042}). The curves for all other values of $m$ are located
  between the $m=1$ and $m\to\infty$ lines in the asymptotic region ($\epsilon_{AA}\gtrsim 15$).} \label{fig1}
\end{figure}

For the ground state, the scaled three-body energies are shown in
figure~\ref{fig1} as function of the scaled two-body energy,
$\epsilon_{AA}$, for
several mass ratios, $m$. The curve for $m=1$ must go through the point
$(1.0,16.52)$ corresponding to three identical bosons. The straight line,
$\epsilon_{3}=\epsilon_{AA}$, is inserted to show that the
three-body system has a stable ground state with binding energy
larger than that of the subsystems. Equivalently, for the binding
energies we have $E_{3}<E_{2}$,
where $E_{2}$ is the smallest of the two-body binding energies $E_{AB}$ and
$E_{AA}$ (below we will use the notation $\epsilon_2=E_2/E_{AB}$ for the corresponding
scaled quantity).
A finite number of stable excited states may also exist
with energies above the $\epsilon_{3}=\epsilon_{AA}$ line.

All curves in figure~\ref{fig1} increase rather steeply from
$\epsilon_{AA}=0$ where $\epsilon_{3}$ must be finite since the binding
energy of the two pairs of particles, $E_{AB}$, is finite. The
behavior of the curves for small $\epsilon_{AA}$ is found to be
\begin{equation}
\epsilon_{3}={\cal F}_0\left( \epsilon_{AA},m\right)
={\cal F}_0(0,m)(1+{\cal G}_{0}(\epsilon_{AA})),\,\,\textrm{for}\,\,
\epsilon_{AA}\rightarrow 0\;,  \label{eq.040}
\end{equation}%
where ${\cal F}_0(0,m)$ is a function of mass, and ${\cal G}_{0}(\epsilon_{AA})$ is
a function of energy with ${\cal G}_{0}(\epsilon_{AA}=0)=0$ and a very
large derivative at $\epsilon_{AA}=0$, like for example
${\cal G}_{0}(x)\propto x^{s}$ with $s$ less than $1$. The limit $\epsilon_{AA} \to 0$
applies to two identical fermions. The singular behavior of the energy in this 
weak-coupling limit
is also found in many-body studies of 2D bosonic systems \cite{schick1971,lieb2001}
and in two-component 2D Fermi gases \cite{bloom1975,randeria1990}. 

The curves in figure~\ref{fig1} all increase almost linearly with
$\epsilon_{AA}$ for large $\epsilon_{AA}$ . The deviation from
linearity is very well approximated by a logarithmic modification
factor, i.e.
\begin{equation}
\epsilon_{3}={\cal F}_0\left( \epsilon_{AA},m\right) \approx
\epsilon_{AA}\ln
\left( \frac{{\cal G}_{1}\left( m\right) }{\epsilon_{AA}}+e\right),\,\, \textrm{for}\,\,
\epsilon_{AA}\rightarrow \infty ,  \label{eq.039}
\end{equation}%
where ${\cal G}_{1}(m)$ is an increasing function of $m$ as seen from the values
given in table~\ref{tab4}. In the extreme limit, we have $\epsilon_3\to \epsilon_{AA}$. 
Here the system is simply that of an $AA$ molecule with binding energy $E_{AA}$ and 
a single $B$ particle with a negligible contribution to the energy.

As the scaled two-body energy $\epsilon_{AA}$ increases, the two
strongly interacting particles contract into one tightly bound dimer entity
for all $m$.
In the limit $\epsilon_{AA}\to\infty $
we are thus left with a two-body problem effectively. It is
known that for any two-body system in 2D we always have a bound
state \cite{artem2011b}, which is the ground state found here. The
three-body binding energy in the limit where $|E_{AA}|\gg |E_{AB}|$,
or $\epsilon_{AA}\gg 1$, is expected to be approximately the
two-body binding energy between the two identical bosons. The other
contributions become of much less importance and give a weak
logarithmic energy dependence. All these features are present in the
asymptotic parametrization in (\ref{eq.039}).

\begin{table}
\centering
\caption{Values of ${\cal G}_{1}(m)$ for $m=0.1$, $1.0$ and $10.0$.}
\begin{tabular}{|c|c|}
\hline
${\cal G}_{1}(0.1)$ & $\simeq 4657$ \\
${\cal G}_{1}(1.0)$ & $\simeq 5751$ \\
${\cal G}_{1}(10.0)$ & $\simeq 12332$ \\ \hline
\end{tabular}
\label{tab4}
\end{table}

\begin{figure}[htb!]
\centering
\epsfig{file=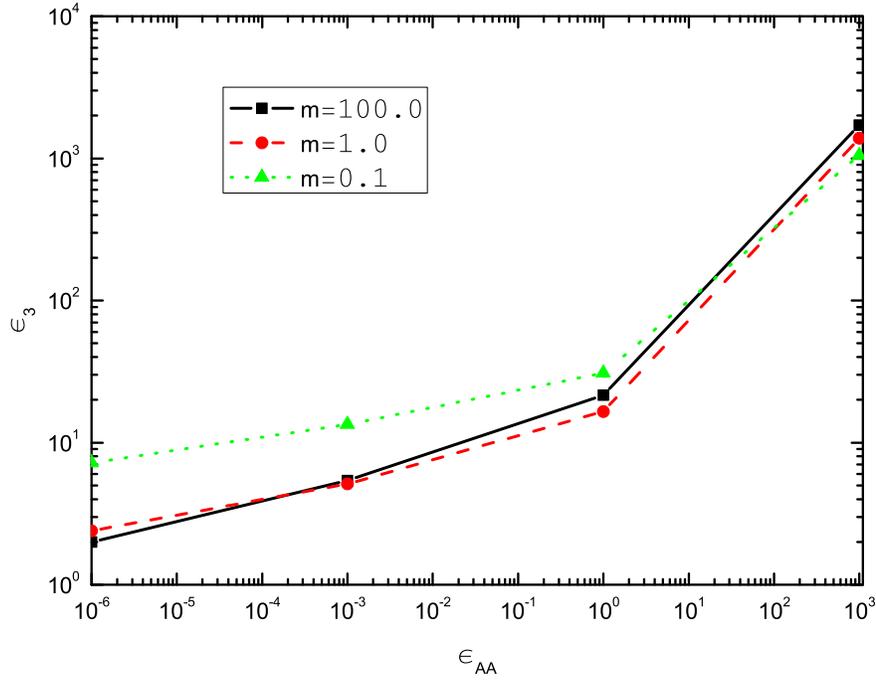,width=0.75\textwidth,angle=0}
\caption{Scaled three-body ground state energy
$\epsilon_3={\cal F}_0(\epsilon_{AA},m)$ for large
variations of $\epsilon_{AA}$ with
$m=0.10$, $m=1.0$, and $m=100$.  } \label{fig3}
\end{figure}

The lines in figure~\ref{fig1} all have qualitatively the same shape.
However, sometimes they cross each other. This behavior is better
appreciated in figure~\ref{fig3} where a much larger variation of
$\epsilon_{AA}$ is shown. When $\epsilon_{AA}\ll 1$, we see that
the three-body energy increases with decreasing mass ratio, whereas
the three-body energy increases with increasing mass ratio for
$\epsilon_{AA}\gg 1$. This feature of figures~\ref{fig1}
and \ref{fig3} is also easily seen in figure~\ref{fig4} below.
The limit of $\epsilon_{AA}\to 0$ on figure~\ref{fig4} agrees well
with the calculation of one fermion and two bosons with resonant 
boson-fermion interaction given in \cite{brodsky2006}.

\begin{figure}[htb!]
\centering
\epsfig{file=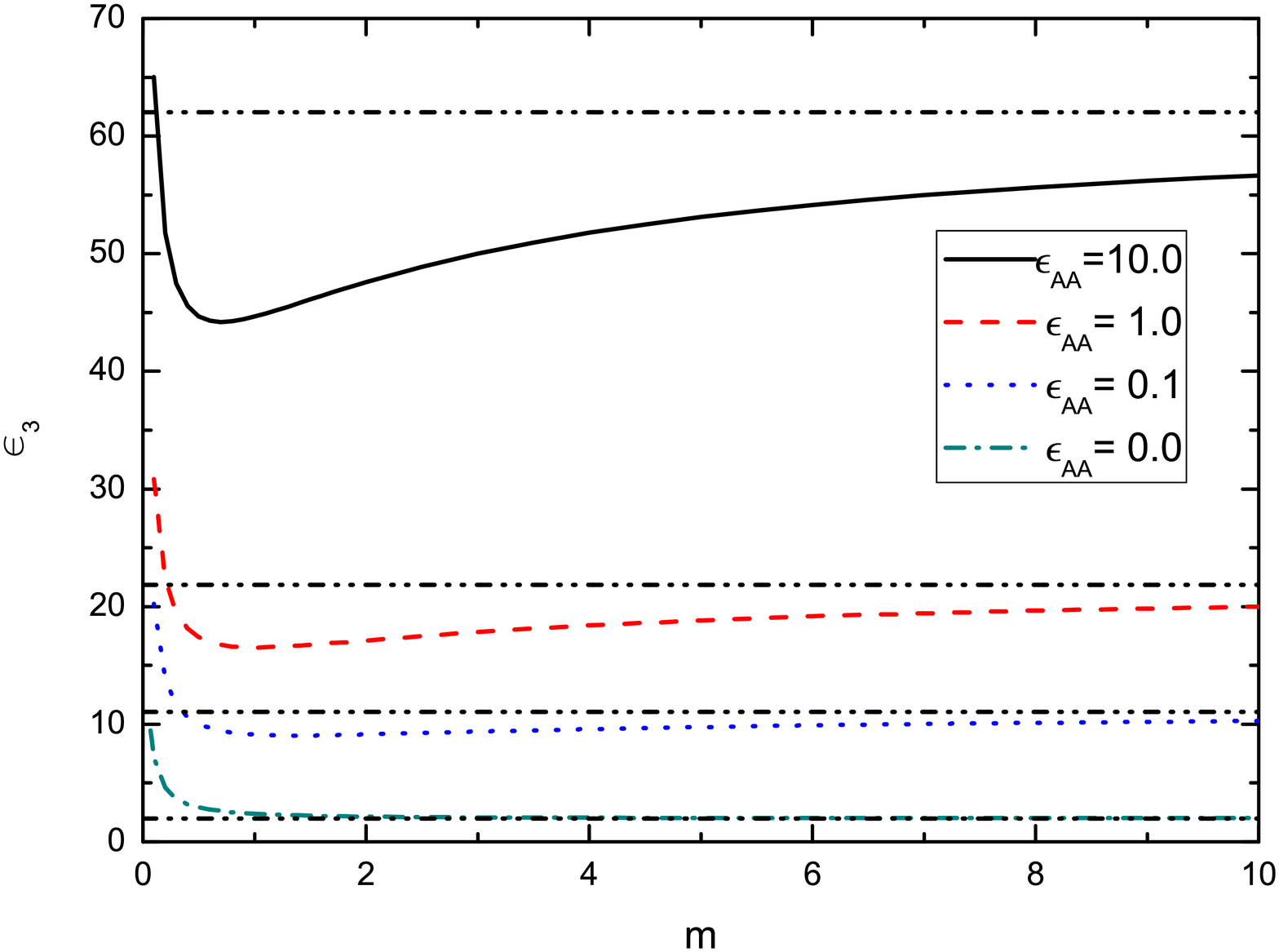,width=0.75\textwidth,angle=0}
\caption{ The scaled three-body ground state energy, $\epsilon_{3}={\cal F}_0(\epsilon_{AA},m)$, as function
  of mass ratio, $ m$, for $ \epsilon_{AA}=0$, $\epsilon_{AA}=0.1$, $\epsilon_{AA}=1$, and
  $\epsilon_{AA}=10$. The dash-dot-dotted lines show the asymptotic large-$m$
  values. } \label{fig4}
\end{figure}

\subsection{Mass dependence}
The mass dependence of ${\cal F}_0$ is shown in figure~\ref{fig4} for the
ground state for several values of
$\epsilon_{AA}$.
When $m$ becomes large, the system consists of one heavy and two light
particles. By rewriting (\ref{eq.026}) and (\ref{eq.027}) for arbitrary $%
\epsilon_{AA}$ in the limit of large $m$, we obtain mass
independent equations, i.e.
\begin{equation}
f_{B}\left( \mathbf{q}\right) =\left[ \pi \ln \left( \sqrt{\frac{\frac{1%
}{4}\mathbf{q}^{2}+\epsilon_{3}}{\epsilon_{AA}}}\right) \right]
^{-1}\int d^{2}k
\frac{f_{A}\left( \mathbf{k}\right) }{\epsilon_{3}+\frac{1}{2}\mathbf{q}^{2}
-\mathbf{k}^{2}-\mathbf{k}\cdot
\mathbf{q}},  \label{eq.031}
\end{equation}
\begin{eqnarray}
f_{A}\left( \mathbf{q}\right) =\left[ 4\pi \ln \left( \sqrt{\frac{1}{2%
}\mathbf{q}^{2}+\epsilon_{3}}\right) \right] ^{-1}
&\left[ \int d^{2}k\frac{f_{B}\left(
\mathbf{k}\right) }{\epsilon_{3}+ \mathbf{q}^{2}-\frac{1}{2}\mathbf{k}^{2}-%
\mathbf{k}\cdot \mathbf{q}}+\right.&\nonumber\\
&\left.\int d^{2}k\frac{f_{A}\left( \mathbf{k}\right) }{\epsilon_{3}+\frac{1}{2}\left(
\mathbf{q}^{2}+\mathbf{k}^{2}\right) }\right]& . \label{eq.032}
\end{eqnarray}
Thus, for large values of $m$ the scaled three-body energy becomes
$m$-independent and approaches an energy-dependent constant as seen
in figure~\ref{fig4}. The limit can be expressed as
\begin{equation}
\epsilon_{3}={\cal F}_0\left( \epsilon_{AA},m\right) \rightarrow
{\cal F}_{0}(\epsilon_{AA},\infty),\,\,\textrm{for}\,\, m\rightarrow \infty,
\label{eq.042}
\end{equation}%
where ${\cal F}_{0}(\epsilon_{AA},\infty)$ is the $m\to\infty$ line shown in figure~\ref{fig1}.
The high-energy behavior is related to (\ref{eq.039}) where
${\cal F}_{0}(\epsilon_{AA},\infty)$ is obtained from ${\cal G}_{1}(m=\infty )$. The
curve ${\cal F}_0(0,\infty)$ approaches the structure of two
mutually non-interacting light particles each interacting with the
same heavy particle. The total energy is therefore a sum of the two
light-heavy particle energies, that is a total energy of $2E_{AB}$, as
seen for large $m$ and $\epsilon_{AA}=0$ in figure~\ref{fig4}. This
can also be considered an accuracy test of the numerical procedure.

Decreasing the mass ratio towards zero leads to an increasing scaled
three-body energy which rises rather quickly when $m$ becomes smaller than
about $0.2$, as seen in figure~\ref{fig4}. In the limit of vanishing $m$,
where the system consists of one light and two heavy particles, all
three-body energies diverge. This is found from the diverging terms in
(\ref{eq.026}) and (\ref{eq.027}), which are compensated by a
diverging three-body energy. Furthermore, infinitely many
bound states simultaneously appear. This behavior can be studied in
the Born-Oppenheimer approximation, where a screened Coulomb-like energy
behavior emerges \cite{bornoppen} in the limit of vanishing
mass ratio $m$. This can be seen in the next section in figure~\ref{fig8}, where
we find that $\epsilon_{3}\propto \frac{1}{m}$ for $m\rightarrow
0$ when $\epsilon_{AA}=0$, that is
\begin{equation}
\epsilon_{3}={\cal F}_n\left( 0,m\right) \rightarrow {\cal B}_{n}\frac{1}{m},\,\,\textrm{for}\,\,
m\rightarrow 0,
\end{equation}
where ${\cal B}_{n}$ is a constant depending on whether ground or excited state is
considered.

Through inspection of figures~\ref{fig1} and \ref{fig4} we realize
that the minimum possible scaled three-body energy is found in the
limits $\epsilon_{AA}\rightarrow 0$ and $m\rightarrow \infty $.
Correspondingly we realize that the maximum scaled three-body energy
is found in the limits $\epsilon_{AA}\gg 1$ and $m\rightarrow
\infty $. Actually the three-body energy diverges for small $m$, so
this maximum is found only if we exclude the very small $m$ region.

The crossing behavior in figures~\ref{fig1}, \ref{fig3}, and \ref{fig4} can be
related to the mass dependence of $\epsilon_{3}$. Let us focus on two
vertical lines in figure~\ref{fig4} for mass ratios $m=0.1$ and $m=10$.
We notice that ${\cal F}_0\left( 0,0.1\right) >{\cal F}_0\left( 0,10\right) $ and
${\cal F}_0\left( 10,0.1\right) <{\cal F}_0\left( 10,10\right) $. Therefore there must
exist a scaled energy, $\epsilon_{AA}$, such that
${\cal F}_0\left( \epsilon_{AA},0.1\right) ={\cal F}_0\left( \epsilon_{AA},10\right)$,
corresponding to the point of crossing. The same procedure can be followed for all the
possible crossing points.

\section{Stability}
Stability of the pure three-body system is determined by the scaled energy $%
\epsilon_{3}$ compared to the thresholds for binding two of the
constituent particles.
This means that if the scaled energy $\epsilon_{3}$ is larger than
all thresholds the three-body system is stable. This suggests to
measure the excess of $\epsilon_{3}$ over the largest scaled
two-body energy, that is $\epsilon_{AA}$ or $1$, when $\epsilon_{AA}$
is larger or smaller than $1$, respectively. For $\epsilon_{AA}\geq 1$
the threshold is increasing with $\epsilon_{AA}$, and
for $\epsilon_{AA}\leq 1$ the threshold is $|E_{AB}|$ which in our units
is equal to 1.

\begin{figure}[htb!]
\centering
\epsfig{file=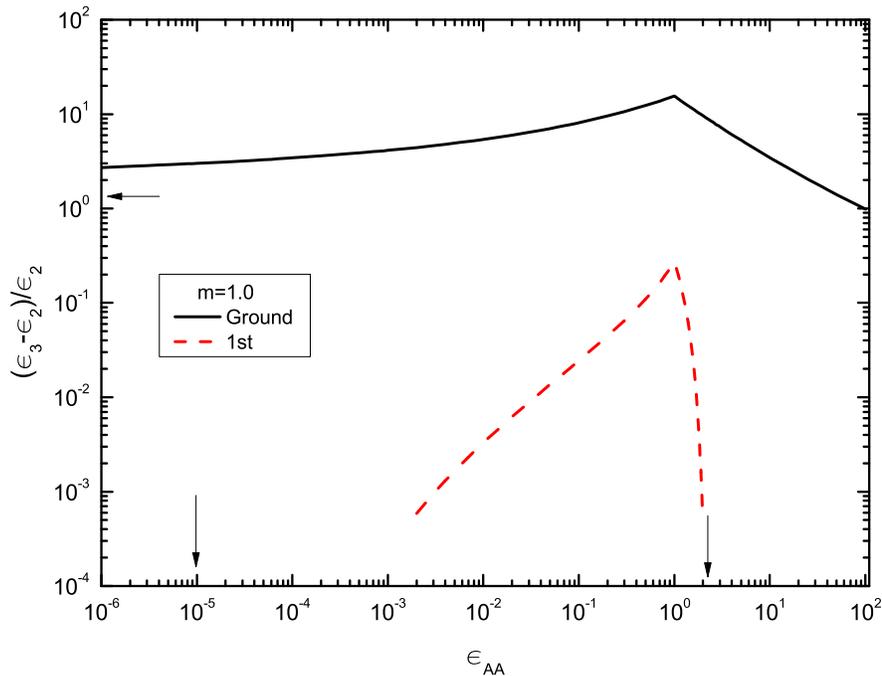,width=0.75\textwidth,angle=0}
\caption{The relative three-body energy, $(\epsilon_3-\epsilon_2)/\epsilon_2$, of ground and first excited states
  for $m=1$ as function of scaled two-body energy $ \epsilon_{AA}$. We
  show the scaled three-body energy relative to the possible thresholds, $\epsilon_{2}$, of
  binding the two two-body subsystems. Here $\epsilon_{2}$ is the largest
  value of $\epsilon_{AA}$ and 1.  For $\epsilon_{AA}\leq 1$ the threshold is
  $1$ and for $\epsilon_{AA}\geq 1$ the threshold is $\epsilon_{AA}$.  The peaks result
  from the different thresholds.  The solid (black) and dashed (red) lines refer
  to ground and excited states, respectively.  The $\downarrow$ arrow
  indicates the points given in table~\ref{tab2}, while the $\leftarrow$
  indicates the $\epsilon_{AA}=0$ limit.} \label{fig7}
\end{figure}

\subsection{Identical masses}
Consider first $m=1$ where we already know that two three-body bound
states exist for three identical bosons. We show in
figure~\ref{fig7} the stability plot for both ground and first excited
states. Stability for both states corresponds to positive
values of $\epsilon_3>\epsilon_{AA}\geq 1$, and values larger than
$1$ for $\epsilon_{AA}\leq 1$. We observe the known ground state
stability for all scaled two-body energies, $\epsilon_{AA}$. The
peaks are an artifact of plotting relative to different
thresholds.

We see in figure~\ref{fig7} how both these states move with respect to
the threshold of stability. The variation is so large that we need
to use log-log scales. For $\epsilon_{AA}=1$ we find the values in
(\ref{eq.030}), that is $16.52$ and $1.27$ for ground and excited
state, respectively. The ground state remains above the thresholds
for all values of $\epsilon_{AA}$. However, the excited state
approaches the threshold of stability for both large and small
$\epsilon_{AA}$, that is one excited state is present
for $m=1$ only when the underlying two-body energies are in the interval, $%
1.00\times 10^{-5}\leq \epsilon_{AA}\leq 2.36$. This is detailed in
table~\ref{tab2} for three different mass ratios.

\begin{figure}[htb!]
\centering
\epsfig{file=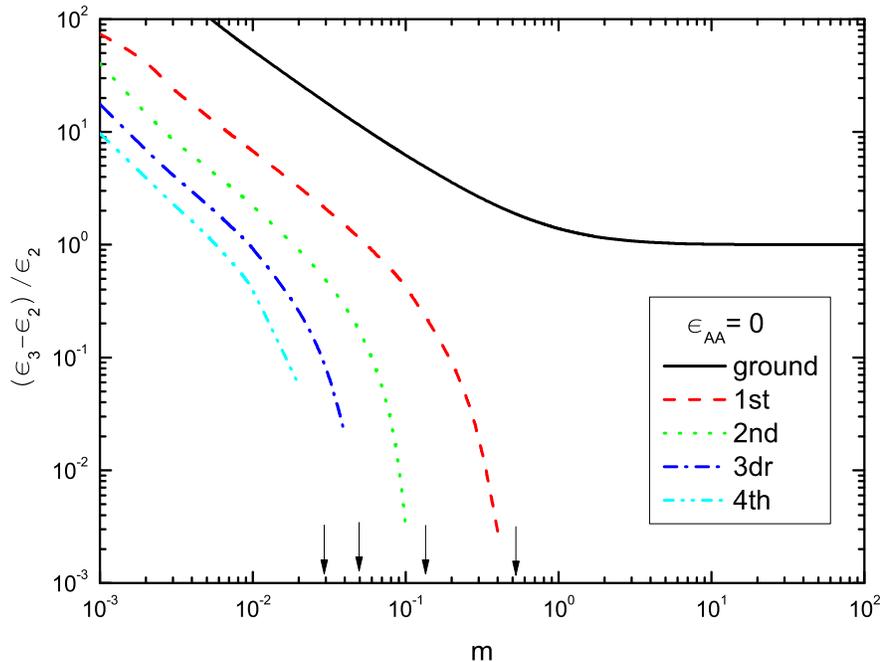,width=0.75\textwidth,angle=0}
\caption{The relative three-body energies, $(\epsilon_3-\epsilon_2)/\epsilon_2$,
for $ \epsilon_{AA}=0$ as function
of $m$. The $\downarrow$ arrows indicate the
$m$-values where the states successively emerge
$\left(\protect\epsilon_{3}=\protect\epsilon_{2}\right)$.  When
$m\to \infty $ we have
$\protect\epsilon_{3}\rightarrow2\protect\epsilon_{2}$.} \label{fig8}
\end{figure}

For non-identical particles with different interactions and masses,
the number of stable bound states may be completely different. The
conditions for existence of excited states can be formulated much
more conveniently for different values of $m$. In general, the
three-body binding energy of any excited state has to be smaller
than the smallest threshold for binding any subsystem. In
figure~\ref{fig8} we see that for $\epsilon_{AA}=0$ (directly applicable for identical fermions)
and $m=1$ only
the ground state is bound. This means that the excited state appears
for $\epsilon_{AA}=0$ when $\epsilon_{3}$ exceeds the threshold of 1,
and ceases to exist for $\epsilon_{3}=\epsilon_{AA}$ when
$\epsilon_{AA}>1$.
These conditions refer to the different thresholds, $\epsilon_{AA}$ and 1,
for $\epsilon_{AA}>1$ and $\epsilon_{AA}<1$, respectively.

We can introduce the threshold condition, $\epsilon_{AA}=0$, in
(\ref{eq.026}) and (\ref{eq.027}). The result is that only one equation 
remains, i.e.
\begin{eqnarray}
f_{A}\left( \mathbf{q}\right) =&\left[ 4\pi \frac{m}{m+1}\ln \left(
\sqrt{\frac{m+2}{2\left( m+1\right) }\mathbf{q}^{2}+\epsilon_{3}}\right)
\right] ^{-1}\times&\nonumber\\
&\int d^{2}k\frac{f_{A}\left(
\mathbf{k}\right) }{\epsilon_{3}+\frac{m+1}{2m}\left(
\mathbf{q}^{2}+\mathbf{k}^{2}\right) +\frac{1}{m}\mathbf{k}\cdot
\mathbf{q}}.& \label{eq.033}
\end{eqnarray}%
This means that the two identical particles in the limit of
vanishing energy, $\epsilon_{AA}=0$, do not feel each other and the distinct
particle only feel the potential separately from one or the other. This 
holds for trivially for identical bosons, and for identical fermions since
we have neglected higher partial waves in the interaction.

\subsection{Excited states}
A survey of the results are shown in figure~\ref{fig8} as function of $m$. The
ground state is always stable and approaches $2$ as $m\rightarrow \infty $.
The excited states appear one after the other as $m$ decreases from $m=1$
towards zero where the system consists of two heavy and one light particle.
From 3D physics we are very familiar with this behavior of a
denser spectrum for such a system. Ultimately, infinitely many stable bound
states exist in the limit of $m=0$.

\begin{table}
\centering
\caption{The mass ratio threshold, $m_{t}$, for appearance of stable
  excited states as the mass ratio decrease towards zero for $\epsilon_{AA}=0$.}
\begin{tabular}{|c|c|}
\hline
1st & $m_{t}\simeq 0.55$ \\
2nd & $m_{t}\simeq 0.12$ \\
3rd & $m_{t}\simeq 0.05$ \\
4th & $m_{t}\simeq 0.03$ \\ \hline
\end{tabular}
\label{tab1}
\end{table}

The log-log plot is necessary in figure~\ref{fig8} for the general overview but does not
expose the precise value of $m$ at the thresholds for the appearance which is defined by a
positive value in figure~\ref{fig8}. These thresholds are instead indicated by
arrows and more precise values of the mass ratios are given in table~\ref{tab1}.
We note in both table and figure how the excited states appear
closer to each other as $m\to 0$.

\begin{figure}[htb!]
\centering
\epsfig{file=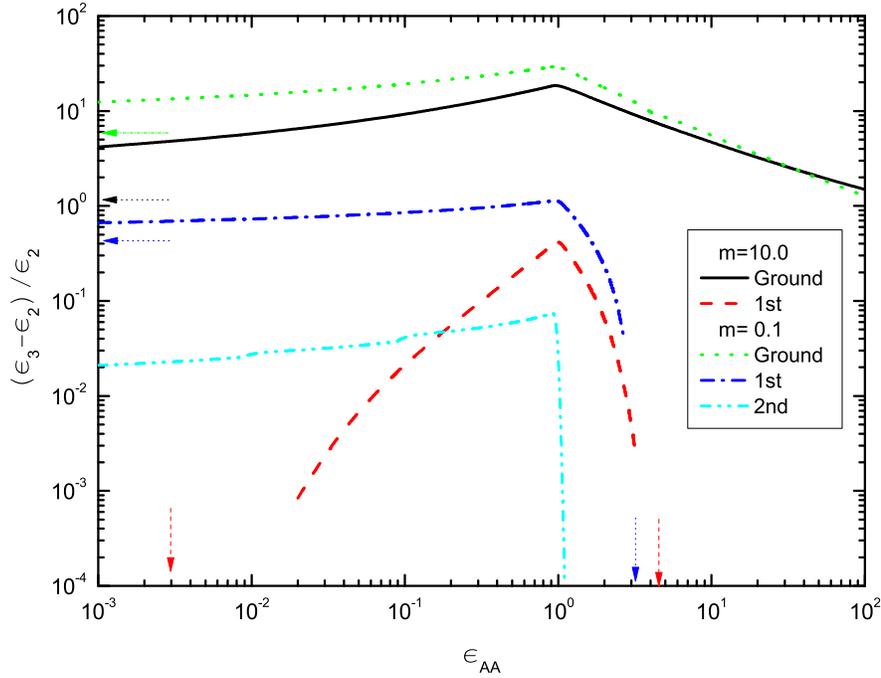,width=0.75\textwidth,angle=0}
\caption{The relative three-body energy, $(\epsilon_3-\epsilon_2)/\epsilon_2$
  of ground and first excited states for
  $m=0.1$ and $m=10$ as function of the scaled two-body energy $\epsilon_{AA}$.
  Here $\epsilon_{2}$ is the largest
  value of $\epsilon_{AA}$ and 1.  For $\epsilon_{AA}\leq 1$ the threshold is $1$
  and for $\epsilon_{AA}\geq 1$ the threshold is $\epsilon_{AA}$.  The peaks result
  from the different thresholds. Solid (black) and dashed (red) lines are ground and first
  excited states, respectively, for $m=10$. Dotted (green), dash-dotted (blue) and dash-double dotted (cyan)
  lines are ground, first, and second states, respectively, for $m=0.1$. The $\downarrow$ arrows mark
  the thresholds given in table~\ref{tab2}.  The $\leftarrow$ arrows
  indicate the $\epsilon_{AA}=0$ limit.}
\label{fig9}
\end{figure}

We now investigate how these results vary as function of $\epsilon_{AA}$
and $m$. We know from figure~\ref{fig8} that for
$m=10$ or $m=0.1$ we have only one or three bound states,
respectively. An interesting question is for which value of
$\epsilon_{AA}$ will the number of bound states change when we keep
$m$ fixed.
The results are shown in figure~\ref{fig9} for $m=10$ and $m=0.1$. Again the
precise thresholds cannot be seen in this figure, and we present therefore
the values in table~\ref{tab2}. We see that for $m=10$ only the ground and
first excited state are stable, while for $m=0.1$ the second excited state
is also found to be stable.

\begin{table}
\centering
\caption{The energy intervals for appearance of stable
  excited states for different mass ratios $m$.}
\begin{tabular}{|c|ccc|}
\hline State & $m$ & $\epsilon_{AA}^{\min }$ &
$\epsilon_{AA}^{\max }$ \\ \hline
& $0.1$ & $0.0$ & $\infty $ \\
gr. & $1$ & $0.0$ & $\infty $ \\
& $10$ & $0.0$ & $\infty $ \\ \hline
& $0.1$ & $0.0$ & $3.76$ \\
1st & $1$ & $1.10\times 10^{-5}$ & $2.36$ \\
& $10$ & $3.00\times 10^{-3}$ & $4.69$ \\ \hline
& $0.1$ & $0.0$ & $1.08$ \\
2nd & $1$ & $-$ & $-$ \\
& $10$ & $-$ & $-$ \\ \hline
\end{tabular}
\label{tab2}
\end{table}

From figures~\ref{fig8} and \ref{fig9}, and from table~\ref{tab2} the question
arises of how low $m$ should be before the next excited state appears. The
condition is that the three-body energy $\epsilon_{3}$ has to exceed both
$\epsilon_{AA}$ and $1$. In other words, the thresholds where the
state begins and ceases to exist are equal. Therefore we search for
the $m$ values
where solutions to (\ref{eq.026}) and (\ref{eq.027}) exist for which
$\epsilon_{3}$ equals the threshold values of $\epsilon_{AA}$ and
$1$. We conclude that both ground and first excited states
always exist for any value of $m$, provided the two-body energy
$\epsilon_{AA}$ assumes an appropriate value depending on $m$,
as we now explain.

\begin{table}
\centering
\caption{The critical mass ratios, $m_c$, above which only $N_c$ bound
  states can exist.}
\begin{tabular}{|cc|}
\hline
$N_{c}$ & $m_{c}$ \\ \hline
2 & 0.22 \\
3 & 0.07 \\
4 & 0.04 \\
5 & 0.02 \\ \hline
\end{tabular}
\label{tab3}
\end{table}

\begin{figure}[htb!]
\centering
\epsfig{file=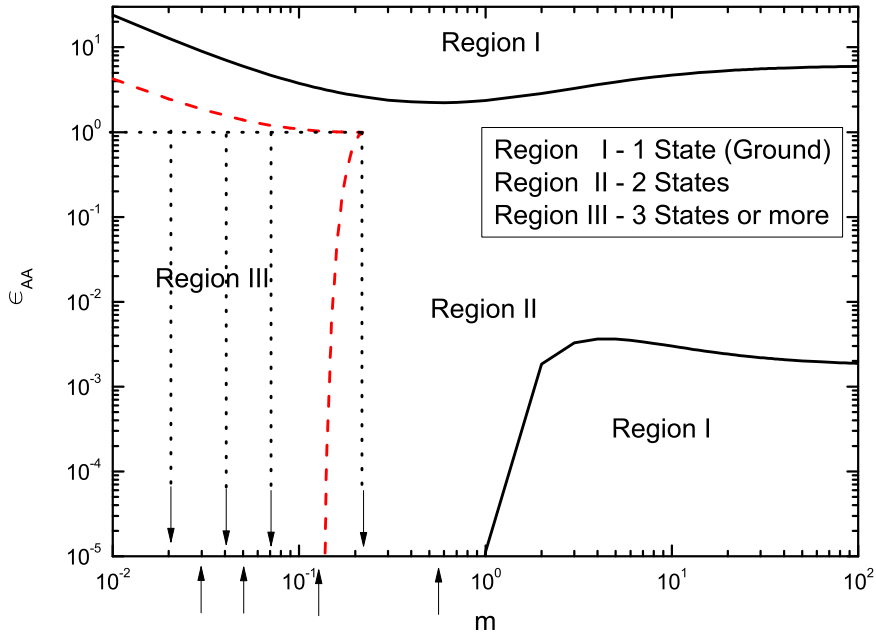,width=0.75\textwidth,angle=0}
\caption{The $m-\epsilon_{AA}$ contour diagram of regions where stable
excited three-body states exist.
  In Region I only the ground state is stable. In
  Region II the ground and the first states are both stable.  In
  Region III three or more states are stable.  The solid (black) and
  dashed (red) lines show the limits where the first and second excited
  states are stable.  The $\downarrow$ arrows mark
  the critical masses, $\left( m_{c}\right) $, for which the next
  excited state appears, table~\ref{tab3}.
  The $\uparrow$ arrows
  indicate the masses for which the next excited state emerges when
  $\epsilon_{2}=0$, table~\ref{tab1}. } \label{fig11}
\end{figure}

A higher number of excited states only exists if the mass ratio $m$ is
sufficiently small. For example if $m\geq 0.22$ only ground and first
excited states can be present. We give in table~\ref{tab3} the critical
masses for appearance of a higher number of bound states. These critical
masses are indicated with arrows in figure~\ref{fig11} where we show the $%
m-\epsilon_{AA}$ phase diagram for the number of stable excited states.
As an example of the use of figure~\ref{fig11}, we can conclude that at
most ground and first excited state are present for $m=1$ and
$m=10$, whereas no more than three bound states exist for $m=0.1$.
It is worth emphasizing that the critical mass for appearance of the
third stable bound state is at the kink of the dashed (red) curve in the
middle of figure~\ref{fig11}. The threshold is not for
$\epsilon_{AA}=0$ as the almost vertical curve otherwise seems to indicate.

\subsection{Realistic systems}
The results in figure~\ref{fig11} may seem strange; the
number of stable bound states decrease when one of the two-body
attractions increase. 
This is similar to the number of Efimov states in a 3D system.
If we start on the side of the resonance where no two-body bound
state is present (negative scattering length, $a<0$, in a typical zero-range 
model, see for instance figure 1 in \cite{ferlaino2010})
then the number of bound states is finite (and can be zero). 
When the attraction to the resonance ($|a|=\infty$) the number of states
is infinite. On the other side of the resonance ($a\to 0^+$) the number of 
bound states will start to decrease again to a finite number.
For the Efimov states the explanation is that, although all
energies decrease with increasing attraction, the three-body states
catch up with the two-body threshold \cite{jen04}.  The result is that
three-body states disappear into the two-body continuum, and the
number of bound states decreases as a consequence. In complete
analogy, the energies of the present excited three-body bound states
also decrease with increasing attraction. However, the two-body
thresholds decrease faster, and the three-body states become
unstable as they merge with the two-body continuum.

In general, figure~\ref{fig11} shows the number of bound states for any
set of parameters $(\epsilon_{AA},m)$. This pair of numbers can be
related to any
set of two-body energies and masses through the scaling relation in
(\ref{eq.023}) and (\ref{eq.025}). Thus for any given point in
figure~\ref{fig11} we find the true three-body binding energy from
figure~\ref{fig1}.
In the Regions I and
II we have one and two stable states, respectively. Region III
collects parameter intervals where more than two stable bound states
are present. This region could be subdivided by a number of curves
similar to the dashed (red) curve. Thresholds for $\epsilon_{AA}=0 $ are
shown but finite values of $\epsilon_{AA}$ extend the mass
regions as the boundary between regions II and III. We expect
that $0.07<m<0.22$ allow up to three while $0.04<m<0.07$ allow up to
four stable bound states. Better values are given in table~\ref{tab3}.

Some commonly used alkali atoms for ultracold atomic experiments
are $^{6}$Li, $^{40}$K, $^{87}$Rb, and $^{133}$Cs.
These mass ratios range from $0.05$ to $22$. The pairwise
interaction is usually tunable by Feshbach resonances. Therefore we have to look for
variations of the dependence on the two-body binding energies. From
figure~\ref{fig11} and table~\ref{tab3} we see that if $m\geq 0.22$ at most
two bound states exist. Whether the first excited state is present
or not depends on $\epsilon_{AA}$. If $m<1$ and $\epsilon_{AA}$ is
larger than about $1$, only the ground state is stable. If $m>1$ and
$\epsilon_{AA}$ is between about $0.001$ and $1$ also the first
excited state is stable.
When $m$ is slightly less than $0.22$ both $1$, $2$, and $3$ stable
states may be present depending on $\epsilon_{AA}$. As $m$
decreases below $0.22$, an increasing number of stable states are
possible when $\epsilon_{AA}$ is sufficiently small as indicated in
figure~\ref{fig11}.

The 2D structures are not yet routinely made but a number of experimental
investigations on identical particles have been reported as discussed in the
introduction. There we also pointed out some modifications expected from the
fact that the experiments are only quasi-2D. If we consider a layered system 
with long-range interactions as shown in figure~\ref{schematic}b)
then particles can be placed in two or three equidistant layers.
For example, if we consider three layers and
place the identical particles in the outer layers and the
distinct particle in the central layer, then the scaled two-body energy
$\epsilon_{AA}$ is about $0.25$ when all particles have the same
mass, see e.g. \cite{jeremy2011,arm11}. Precisely two bound states are present
in the universal regime for this setup.
This still holds when the mass ratio is larger than $1$. However,
when the mass ratio decreases, more excited states may be present.

The physics of highly polarized Fermi gases is also interesting in relation
to the present study. The problem of a single impurity interacting
with a Fermi sea of particles in 2D has generated considerable theoretical interest 
recently \cite{zollner2011,parish2011,klawunn2011}. Furthermore, fermionic 
impurities in Bose-Einstein condensates have recently been studied in 
experimentally in optical lattices (although so far only in 3D)
\cite{will2011}. Theoretical studies 
of these kinds of setups usually focus mostly on the influence of 
two-body bound states. However, we expect that interesting spectra of
three-body states can occur in these systems and it will be interesting
to study how a many-body background such as a Fermi sea or a condensate
can affect the properties of three-body states.

\section{Summary and conclusions}
In the present paper we investigate the three-body problem in two dimensions.
The aim is to extract universal properties where
any potential with similar (observable) constraints is able to
describe the model-independent results.
This only occurs when the
properties hinges on large-distance behaviour where the details of the
basic two-body ingredients are unimportant. Zero-range models are
then suitable since all properties are determined at distances outside
the potential. We therefore employ the established method of solving the
momentum-space Faddeev equations with zero-range interactions.

We focus for simplicity on a three-body system with two identical
and one distinct particle. Three different particles would be
essentially as easy to solve but the number of free parameters would
be doubled and the results much harder to display and digest. We
leave this generalization for a separate future investigation.
We first establish that the three-body energy in units of one of the
two-body binding energies $E_{AB}$ must be a function of only two
parameters; the ratio of the other two-body binding
energy $E_{AA}$ to $E_{AB}$, and the mass ratio $m=m_B/m_A$.
We investigate this two-parameter problem as function of the
reduced energy and mass parameters. When the identical particles are
fermions their binding energy is zero in the zero-range model,
i.e. $E_{AA}=0$.

The starting point is three identical bosons where the energies of
the two stable bound states are well-known. The scaled three-body
energy is calculated as function of the scaled two-body energy for
fixed mass ratio and vice versa.
The energy dependence is the stronger than the mass dependence and always
monotonically increasing, and we establish logarithmic dependence of
the scaled three-body energy for
large two-body energy.
The mass dependence for fixed scaled two-body
energy is in general smaller but non-monotonic. It is divergently
increasing when the mass ratio approaches zero, while an energy-dependent
and mass-independent constant is approached for large mass
ratios.

A number of excited stable bound states can exist. In general when
two equal masses are not large compared to the third, ground and
first excited states exist and are stable. These two lowest states
are always possible for all mass ratios but, whenever the mass ratio
is outside the interval $0.22 < m <1 $, the first excited state only
occurs for a rather narrow band of two-body energies. As the two
equal masses become heavier the number of stable bound states
increase towards infinity.
This happens when the
mass ratio approaches zero and corresponds to the Born-Oppenheimer limit.
We derive a number of threshold values of masses
and energies for appearance of these stable excited states. Finally,
we provide an energy-mass phase-diagram of regions for occurrence of
specific numbers of excited states and discuss applications for
realistic systems.

In conclusion, we have established the two-dimensional universal energy
relations for a three-body system of two identical and one distinct
particle. These investigations are of interest as properties of basic
quantum mechanical problems extended to two dimensions where the
behavior differ qualitatively from that of three dimensions. We furthermore
expect that our results will become directly relevant in the
topical studies of two-dimensional systems in cold atomic gases.

\ack
This work was partly support by funds
provided by Funda\c c\~ao de Amparo \`a Pesquisa do Estado de S\~ao
Paulo and Conselho Nacional de Desenvolvimento Cient\'\i fico e
Tecnol\'ogico (Brazil). FFB, MTY and TF thanks the hospitality of
the Department of Physics and Astronomy of Aarhus University during
their visits.

\end{document}